\documentclass[conference]{IEEEtran}
\IEEEoverridecommandlockouts
\usepackage{cite}
\usepackage{amsmath,amssymb,amsfonts}
\usepackage{algorithmic}
\usepackage{graphicx}
\usepackage{textcomp}
\usepackage{svg}
\usepackage{braket}
\usepackage{multicol}
\usepackage{multirow}
\usepackage{makecell}
\usepackage{xcolor}
\usepackage{caption}
\usepackage{setspace}
\usepackage{tikzpagenodes}
\usepackage{dblfloatfix}

\usepackage{fancyhdr}
\fancypagestyle{firstpage}{
   \fancyhf{} 
   \fancyhead[C]{To appear at 2024 International Joint Conference on Neural Networks (IJCNN) Proceedings} 
}

\def\BibTeX{{\rm B\kern-.05em{\sc i\kern-.025em b}\kern-.08em
    T\kern-.1667em\lower.7ex\hbox{E}\kern-.125emX}}
\begin{document}

\title{Investigating the Effect of Noise on the Training Performance of Hybrid Quantum Neural Networks \\
}


\author{\IEEEauthorblockN{Muhammad Kashif, Emman Sychiuco,  Muhammad Shafique}
\IEEEauthorblockA{eBrain Lab, Division of Engineering,\\ 
Center for Quantum and Topological Systems, NYUAD Research Institute\\
New York University Abu Dhabi, PO Box 129188, Abu Dhabi, UAE\\
Emails: \{jes9843, muhammadkashif, muhammad.shafique\}@nyu.edu}
}


\maketitle
\thispagestyle{firstpage}

\begin{abstract}
The Hybrid Quantum Neural Networks (HyQNNs) hold great promise for various quantum machine learning tasks, but their performance can be significantly affected by the quantum noise in NISQ devices. 
In this paper, we comprehensively analyze the influence of different quantum noise gates, including Phase Flip, Bit Flip, Phase Damping, Amplitude Damping, and the Depolarizing Channel, on the performance of HyQNNs. Our results reveal distinct and significant effects on HyQNNs training and validation accuracies across different probabilities of noise. For instance, 
the Phase Flip gate introduces phase errors, and we observe that HyQNNs exhibit resilience at higher probability ($p=1.0$), adapting effectively to consistent noise patterns, whereas at intermediate probabilities, the performance declines.
Bit Flip errors, represented by the Pauli X gate, impact HyQNNs in a similar way to that Phase Flip error gate. The HyQNNs, can adapt such kind of errors at maximum probability ($p=1.0$).
Unlike Phase and Bit Flip error gates, Phase Damping and Amplitude Damping gates disrupt quantum information, with HyQNNs demonstrating resilience at lower probabilities but facing challenges at higher probabilities. 
Amplitude Damping, in particular, poses efficiency and accuracy issues at higher probabilities, however with lowest probability ($p=0.1$),it has the least effect, i.e., HyQNNs, although not very effectively, but still tends to learn.
The Depolarizing Channel proves most detrimental to HyQNNs performance, with limited or no training improvements. There was no training potential observed regardless of the probability of this noise gate.
These findings underscore the critical need for advanced quantum error mitigation and resilience strategies in the design and training of HyQNNs, especially in environments prone to depolarizing noise. This paper quantitatively investigate that understanding the impact of quantum noise gates is essential for harnessing the full potential of quantum computing in practical applications.

\end{abstract}




\section{Introduction}
The Noisy Intermediate-Scale Quantum (NISQ) devices represent a significant advancement in the field of quantum computing. 
NISQ devices do not employ full-scale error correction, largely due to their limited qubit count and technological constraints, leading to inherent noise in quantum computations. 
The concept of NISQ devices was first introduced by John Preskill in 2018, emphasizing the potential of these systems for solving specific problems faster than classical computers, despite their imperfections \cite{Preskill:2018}. 
NISQ devices are crucial for current research and applications in quantum computing, as they provide a platform for developing and testing quantum algorithms, error mitigation techniques, and exploring quantum supremacy for certain tasks \cite{Wu:2021,Madsen:2022}. The NISQ era is a transitional phase, expected to evolve into more advanced quantum computing technologies as error correction and qubit control improve over time \cite{lau:2022}.

Despite their inherent noise and operational constraints, NISQ devices hold immense potential for exploring various classical applications in the post-quantum era. One notable application is \emph{Quantum Machine Learning} (QML), an emergent field resulting from the conceptual integration of classical machine learning  with quantum computing principles \cite{Cerezo:2021a}. QML aims to leverage the unique capabilities of quantum systems such as superposition and entanglement to enhance or potentially revolutionize traditional machine learning algorithms. 
In context of NISQ devices, the anticipation here is that even with these limitations, NISQ devices will be capable of executing certain QML algorithms faster or more efficiently than their classical counterparts\cite{Huang:2022, Huang:2022a}, providing early insights into the potential transformative impact of quantum computing on machine learning.

\begin{figure}[htbp]
    \centering
    \includegraphics[scale=0.45]{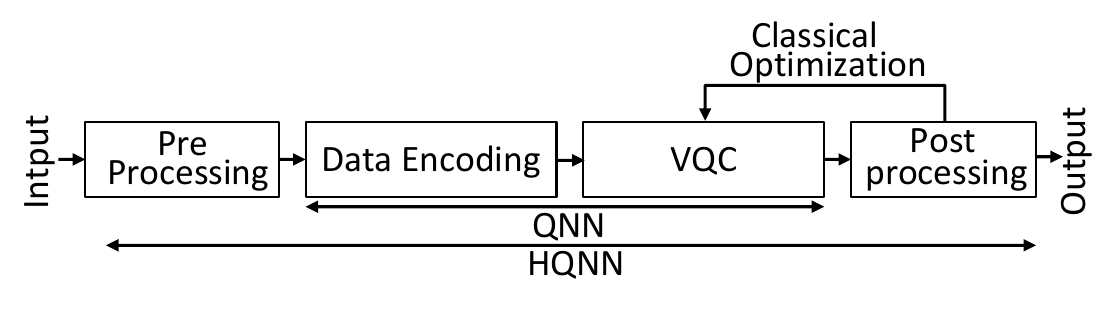}
    \vspace{-15pt}
    \caption{\footnotesize A Typical Illustration of QNN and HyQNNs architecture}
    \label{fig:fig1}
\end{figure}

Due to the limitation of the NISQ devices, a hybrid approach that combines classical and quantum processing, compatible with NISQ devices, has emerged as a leading candidate for exploring quantum advantages in various QML applications \cite{Bergholm:2018}.
To this end, variational quantum circuits (VQCs), have emerged as a promising quantum analog to artificial neurons in classical neural networks \cite{Kashif:2023alleviating} The significance of VQCs in different contexts is underscored by extensive research, as detailed in \cite{Cerezo:2021a, Abbas:2021}. 
VQCs are quantum circuits that can be optimized using classical methods and are notably robust against errors typical in NISQ devices \cite{schuld:2018,kashif_unified}.
When VQCs are coupled with data embedding techniques, they are often referred to as quantum neural networks (QNNs). QNNs have been proposed for a variety of applications, including transfer learning \cite{Mari:2020}, generative modeling \cite{Coyle:2020}, and classification tasks \cite{Farhi:2018,kashif:2021}. Analogous to classical neural networks, where classical parameters are optimized during training, QNNs require the optimization of quantum parameters also \cite{Cerezo:2021a}. The HyQNNs typically have the same architecture as QNN as shown in Fig. \ref{fig:fig1}. However, they are called \emph{Hybrid} because of the use of classical optimization routines and classical pre- and post-processing layers \cite{Kashif:2023alleviating}.
%

\begin{figure*}[htbp]
\centering
    \includegraphics[scale=0.52]{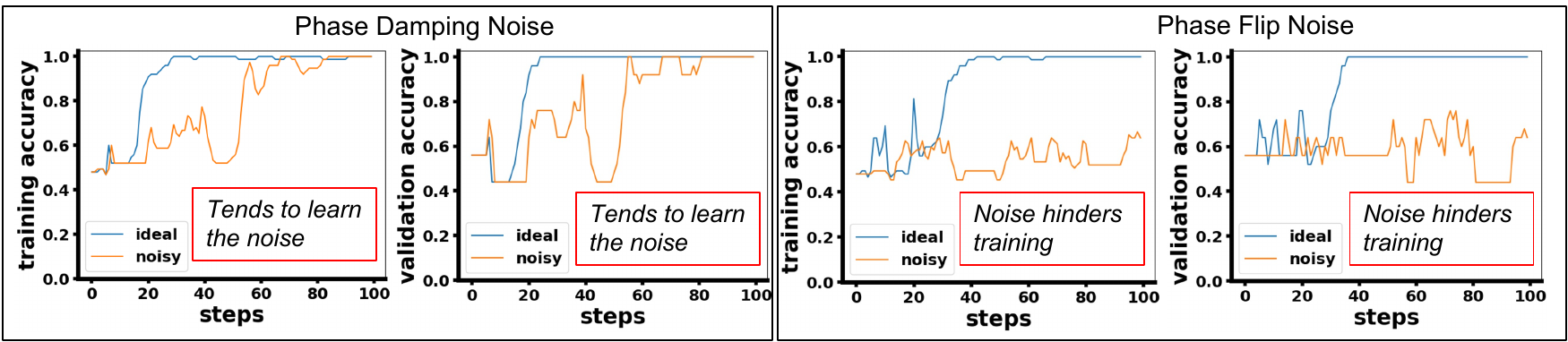}
    \vspace{-7.5pt}
    \caption{\footnotesize  Motivational Analysis. The varying effects of different noise gates, despite having identical probabilities, on the training of Hybrid Quantum Neural Networks (HyQNNs) highlight the necessity for a comprehensive analysis of how various noise gates, each with their own probability ranges, influence HyQNNs training.}
    \label{fig:motiv_analysis}
\end{figure*}

While many works study the trainability and underlying challenges in QNNs \cite{McClean:2018,kashif:2023params, kashif2024resqnets,kashif2024resqunnstowards}, most of them consider ideal scenarios and does not take into the effect of hardware noise, which is an intrinsic characteristic of NISQ devices. 
As we delve deeper into the NISQ era, it is imperative to investigate the limitations imposed by noise while simultaneously exploring innovative strategies to exploit these quantum systems effectively.


\vspace{-7pt}




\subsection{Related Work}
A detailed analysis of how noise impacts QNNs is presented in \cite{escudero:2023}. The focus is on the Mottonen state preparation algorithm and how various noise models affect the degradation of quantum states in QNNs. The study employs the MNIST dataset to test the inference capacity of trained QML models under noisy conditions. It was observed that the noise perturbs the state of quantum systems, and high noise errors in gates or readout result in a faulty distribution of the state. \textit{However, the focus is on the pretrained QNNs whereas in this paper, we train the QNNs with various noise models and analyze if the QNNs can learn the noise over time and make themselves noise aware. } 

The effect of depolarizing noise in QNNs, particularly focusing on adversarial accuracy and defense strategies is presented in \cite{winderl:2023}. The paper discusses the robustness of these networks against various types of attacks, providing insights into the impacts of depolarizing noise on QNN security and reliability. \textit{This research \cite{winderl:2023} only focuses on depolarizing noise type whereas in this paper, we study the effect of a wide range of noise types in NISQ devices on the training performance of HyQNNs. }

An investigation of how noise influences the phenomenon of overparametrization in QNNs is discussed in \cite{garciamartin:2023}, where the quantum Fisher information matrix is used to understand the effects of noise on QNNs. It was found that the presence of quantum noise can increase the rank of the QFIM and decrease the overall magnitude of its eigenvalues, impacting the network's overparametrization. \textit{Although this study provides important insights in from the overparameterization perspective in the presence of noise, it does not discuss that whether or not the QNNs can learn the noise, if trained with high probability errors. }

These studies collectively contribute significant insights into the effects of various noise types on the performance and robustness of QNNs. 
However, we believe that a comprehensive understanding of the influence of various noise types in NISQ devices, particularly from a training perspective in QNNs, remains an under-explored area. 

\textbf{Motivational Analysis:} Furthermore, as a motivational case study, we train HyQNNs with 2 qubit layers on Iris dataset (2 features) with two different noise gates namely; Phase Damping and Phase Flip, the results of which are shown in Fig \ref{fig:motiv_analysis}. For the motivational analysis, the classical input features were encoded into the rotation angles of $RX$ gate. Moreover, an arbitrary parameterized unitary gate was applied on each qubit. Both the qubits were entangled using $CNOT$ gate. The training was performed for a total of 100 steps. Both the noise gates have the occurence probability of 0.1. We observe that different noise gates with same probabilities have distinct affects on model's performance. For instance, the HyQNNs tends to learn the noise patterns in case of phase damping whereas in case of Phase Flip noise, HyQNNs struggles and does not train effectively.

This necessitates further investigation from multiple perspectives to fully understand the implications of different types of quantum noises with a range of different probabilities on the learning capabilities of HyQNNs in the NISQ era.

\vspace{-5pt}

\subsection{Our Contributions}
The key contributions of our paper are summarized below:
\begin{itemize}
    \item \textbf{Comprehensive Analysis of Quantum Noise Gates on HyQNNs:}
    We have conducted an in-depth quantitative analysis on how various quantum noise gates (Phase Flip, Bit Flip, Phase Damping, Amplitude Damping, and the Depolarizing Channel) affect the performance of Hybrid Quantum Neural Networks (HyQNNs).

    \item \textbf{Identification of Distinct Noise Gate Effects:}
    We demonstrate that different quantum noise gates have unique and significant impacts on the training and validation accuracy of HyQNNs, thus providing insights into how each type of quantum noise influences HyQNN's performance.

    \item \textbf{Resilience of HyQNNs to Phase and Bit Flip Errors:}
    One of the key findings of our work is that HyQNNs show resilience to Phase and Bit Flip errors at higher probability (typically $1.0$), effectively adapting to consistent noise patterns. However, at intermediate probabilities ($< 1$), the performance significantly declines.

    \item \textbf{Impact of Phase and Amplitude Damping Gates:}
    We find that Phase Damping and Amplitude Damping gates disrupt quantum information processing in HyQNNs.
    Unlike Phase and Bit Flip errors, HyQNNs show resilience at lower probabilities for Phase and Amplitude Damping erros, but encounter significant challenges at higher probabilities.

    \item \textbf{Effects of the Depolarizing Channel:}
    Moreover, we identify that the Depolarizing Channel has the most detrimental affect on HyQNN's performance, as the HyQNNs shows almost no improvements in training under the influence of the Depolarizing Channel regardless of the error probability.

    \item \textbf{Highlighting the Need for Quantum Error Mitigation Strategies:}
    Based on our findings, we motivate and emphasize the critical necessity of developing advanced quantum error correction for HyQNNs particularly in environments prone to depolarizing noise and error mitigation for other noise types studied in this paper.
\end{itemize}




\vspace{-7pt}
\subsection{Organization}
Section \ref{sec:background} presents a comprehensive overview of HyQNNs and the specific noise gates used in this paper. Section \ref{sec:methodology} presents the methodology adopted for our research, detailing the theoretical and practical approaches. Experimental setups, including the configurations and parameters used, are elaborated in Section \ref{sec:exp_setup}. The findings from these experiments are thoroughly analyzed and discussed in Section \ref{sec:results}. Finally, Section \ref{sec:conclusion} concludes the paper.

\vspace{-7pt}
\section{Background} \label{sec:background}
\vspace{-5pt}
\subsection{Hybrid Quantum Neural Networks}
\begin{figure}[htbp]
    \hspace{-12pt}
    \includegraphics[scale=0.37]{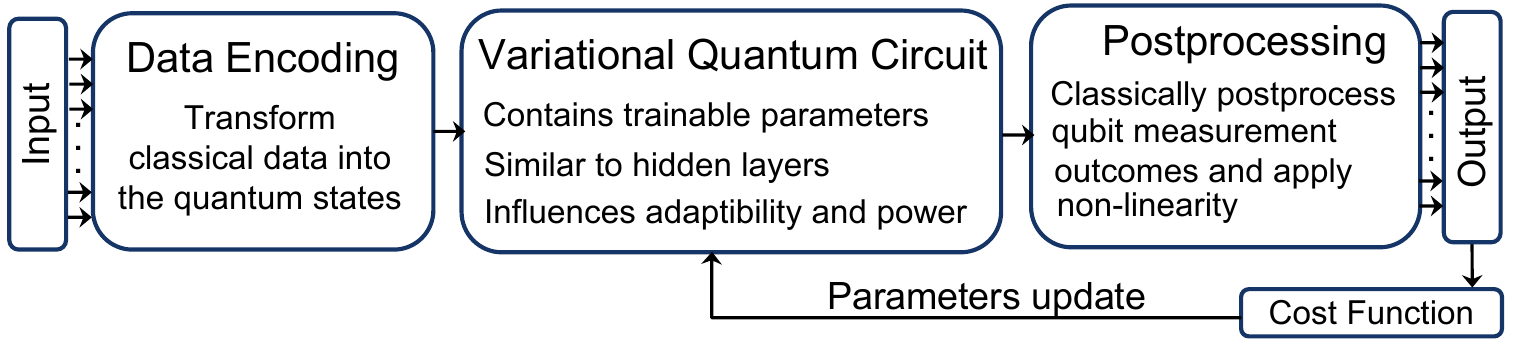}
    \vspace{-10pt}
    \caption{\footnotesize General Architecture of Hybrid Quantum Neural Networks. HyQNNs generally work in three steps: (1)Data Encoding from classical to quantum, (2) Variational circuit training and (3) classical post ptocessing of results obtained from variational circuits)}
    \label{fig:HyQNNs_arch}
\end{figure}

A Hybrid Quantum Neural Network (HyQNN) combines traditional artificial neural network structures with quantum computing principles. Utilizing qubits for information processing, HyQNNs incorporate quantum gates to manipulate these qubit states, thus harnessing quantum parallelism and entanglement. This integration enables HyQNNs to process multiple solutions simultaneously, which could provide computational advantages in specific tasks. Figure \ref{fig:HyQNNs_arch} illustrates the general architecture of HyQNNs, with a subsequent discussion on its various operational stages.

\vspace{-5pt}
\paragraph{Data Encoding}
The initial phase in the operation of an HyQNNs is the encoding of classical data into quantum states. This encoding process is essential as it transforms classical data into a format amenable to quantum computation, thereby facilitating the subsequent quantum information processing.
We have used angle encoding in this paper, therefore, we provide an overview of this technique, however, for more data encoding techniques, please refer to \cite{schuld:2018}.

Angle encoding, encodes the data into the rotation angles of qubits \cite{LaRose:2020}, and is conjectured to be more advantageous for HyQNN applications. Angle encoding typically assigns a single input feature per qubit, thus allowing an $n$-qubit system to encode $n$ input features. A typical representation of angle encoding is given in Eq. \ref{eq1}. 
\vspace{-6pt}
\begin{equation}\label{eq1}
    \footnotesize S_{{x}_{j}} = \bigotimes_{i=1}^{N} U_i \hspace{0.2cm} where \hspace{0.2cm}U_i :=
    \begin{bmatrix}
 \cos(x_j^{(i)})  &  -\sin(x_j^{(i)})\\ 
 \sin(x_j^{(i)})   &\cos(x_j^{(i)}) 
 \end{bmatrix}
\end{equation}

\paragraph{Variational Quantum Circuit}

At the heart of HyQNNs is the Variational Quantum Circuit (VQC), which functions similarly to the hidden layers in classical neural networks. VQCs, containing trainable parameters, perform quantum computations on input data, with their structure crucial to the network's expressive power and adaptability. 
After these computations, quantum states are measured to yield classical outcomes.
In VQCs, the output, often represented as the expectation value of observables, can be formulated as a 'quantum function' ($f(\theta)$), parameterized by a set of parameters $\theta = \theta_1, \theta_2, \ldots$.
An interesting aspect of VQCs is that the partial derivatives of ($f(\theta)$) can be represented as a linear combination of other quantum functions. 
The partial derivative of a VQC parameters can be calculated using the parameter shift rule \cite{Mitarai:2018}, represented in Eq. \ref{eq_param} where $s$ is the shift argument.
\vspace{-5pt}

\begin{equation} \label{eq_param}
    \footnotesize \frac{df}{d\theta_i} = \frac{f(\theta_i+s) - f(\theta_i-s)}{2}
\end{equation}
\vspace{-8pt}

\paragraph{Postprocessing}
The qubit measurement results from VQC are classical in nature and can be processed classically.
The post-processing classical layer interprets and refines the probabilistic outcomes obtained from quantum measurements, generating final, deterministic results for practical use. It acts as a bridge between quantum computations and actionable information in classical applications. 
A classical neuron layer is typically used for postprocessing, and involves the computation of a cost or loss function based on the quantum measurements \cite{Kashif:2023alleviating}. 
Classical optimization algorithms, such as gradient descent, are then employed to iteratively adjust the parameters of the VQC, minimizing the cost function until the solution is reached.

\begin{figure*}[htbp]
    \centering
    \includegraphics[scale=0.6]{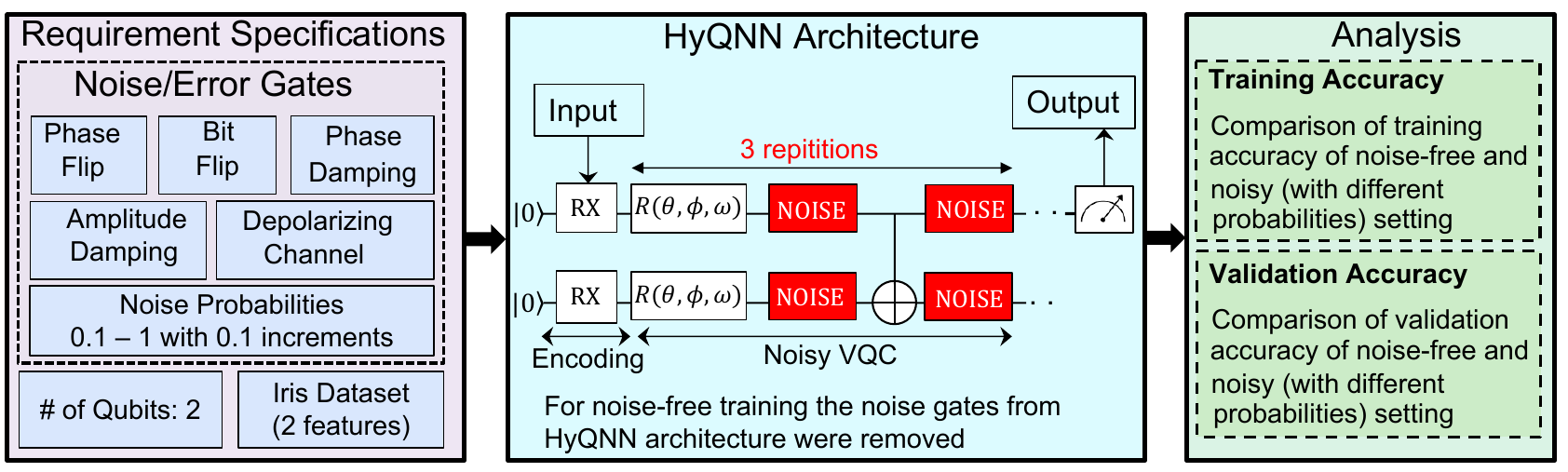}
    \vspace{-8pt}
    \caption{\footnotesize Detailed methodology highlighting keys steps of our analysis. We investigate the influence of different types of quantum noises with different occurrence probabilities on the training of HyQNNs. A simple 2-qubit architecture is used with a depth of three layers. Noise is systematically introduced in each layer, and the performance is compared against a noise-free baseline. Training and validation accuracies are used as the primary metrics to evaluate the impact of noise on the HyQNNs' training performance.}
    \label{fig:methodology}
\end{figure*}

\subsection{Quantum Noise and Error gates}
Quantum noise refers to the uncertainty and fluctuations in quantum systems, stemming from factors like environmental interference and imprecise control over qubits. Unlike classical systems, where noise is generally deterministic, quantum noise is probabilistic due to the principles of quantum mechanics \cite{Yuxuan:2021}. 
%
Error gates are introduced to simulate quantum noise within a quantum circuit to intentionally introduce errors, mimicking the unpredictable effects of quantum noise in NISQ devices. This paper uses a variety of error gates from pennylane’s extensive library \cite{Bergholm:2018}, which we discuss in more detail below.
\paragraph{Phase Flip}

A phase flip is a quantum operation that alters the phase of a quantum state without changing its probability amplitudes. In quantum computing, it is represented as a unitary phase flip gate or Z gate. 
The phase flip gate, when applied to a qubit, leaves the $\ket{0}$ state unchanged and multiplies the $\ket{1}$ state by $-1$. This changes the phase of the $\ket{1}$ state by 
$\pi$ radians but does not affect the probability of measuring either state. Mathematically, phase flip error gate can be expressed as the following Kraus Matrices, where $p \in [0,1]$  is the probability of a phase flip (Pauli Z) error.
\vspace{-5pt}
  $$\footnotesize K_0 = \sqrt{1-p} \begin{pmatrix} 1 & 0\\ 0 & 1 \end{pmatrix}, K_1 = \sqrt{p} \begin{pmatrix} 1 & 0\\ 0 & -1 \end{pmatrix}$$ 

\vspace{-5pt}


\paragraph{Bit Flip}

A bit flip error in quantum computing is analogous to a classical bit flip. In classical computing, a bit flip error refers to the unwanted switching of a bit's state from 0 to 1, or vice versa. Similarly, in quantum computing, a bit flip error is an error that causes a qubit in the state $\ket{0}$ to flip to $\ket{1}$ and vice versa.
Bit flip error is a Pauli X error on a qubit. This can be expressed using the following Kraus Matrices, where $p \in [0,1]$  is the probability of a bit flip (Pauli X) error.
\vspace{-10pt}

    $$\footnotesize K_0 = \sqrt{1-p} \begin{pmatrix} 1 & 0\\ 0 & 1 \end{pmatrix}, K_1 = \sqrt{p} \begin{pmatrix} 0 & 1\\ 1 & 0 \end{pmatrix}$$

\vspace{-2pt}

\paragraph{Phase Damping}

Phase damping is a type of quantum noise that affects the phase information of a qubit while leaving its probability amplitudes unchanged. This phenomenon is distinct from other types of quantum errors like bit flips or phase flips, as it does not directly change the state of the qubit from $\ket{0}$ to $\ket{1}$ or vice versa. Instead, phase damping leads to a loss of quantum coherence by gradually diminishing the off-diagonal elements of a qubit's density matrix.
This error can be modeled using the following Kraus Matrices where $\gamma \in [0,1]$ is the phase damping probability.
\vspace{-4pt}
    $$\footnotesize K_0 = \begin{pmatrix} 1 & 0\\ 0 & \sqrt{1-\gamma} \end{pmatrix}, K_1 =  \begin{pmatrix} 0 & 0\\ 0 & \sqrt{\gamma} \end{pmatrix}$$ 
  
\vspace{-4pt}

\paragraph{Amplitude Damping}

Amplitude damping is a type of quantum noise that represents the loss of energy from a quantum system. This process is similar to the natural decay of an excited quantum state to a lower energy state, often modeled as the interaction of a qubit with an external environment, leading to the qubit losing its excitation over time.
Amplitude damping can be viewed as the process where a qubit in the excited state $\ket{1}$ has a probability of decaying to the ground state $\ket{0}$. This is different from phase damping, which affects the relative phase between the quantum states but does not cause a transition between states.
Amplitude damping can be modeled by the following Kraus Matrices, where $\gamma \in [0,1]$ is the amplitude damping probability.

\vspace{-5pt}
$$\footnotesize K_0 = \begin{pmatrix} 1 & 0\\ 0 & \sqrt{1-\gamma} \end{pmatrix}, K_1 =  \begin{pmatrix} 0 & \sqrt{\gamma}\\ 0 & 0 \end{pmatrix}$$ 

\vspace{-4pt}

\paragraph{Depolarizing channel}
The depolarizing channel is a type of quantum noise model, which represents a process, where a qubit loses its information without favoring any particular basis. The depolarizing channel is an important noise model  where the probability of error is the same in all directions of the Bloch sphere.

In the depolarizing channel noise model, a qubit in any given state $\rho$ is replaced with maximally mixed state
$I/2$ (where I is the identity matrix) with some probability $p$. The maximally mixed state represents a state of complete uncertainty, with equal probabilities of being in $\ket{0}$ and $\ket{1}$ states. 
The effect of the depolarizing channel is to reduce the purity of the qubit state, effectively "smearing" its position on the Bloch sphere towards the center, which corresponds to the maximally mixed state. As $p$ increases, the state becomes more mixed, losing quantum information and coherence.
A depolarizing channel can mathematically be modeled by the following Kraus matrices, where $p \in [0,1]$ is the depolarization probability and is equally divided in the application of all Pauli operations.
\vspace{-8pt}
$$\footnotesize K_0 = \sqrt{1-p} \begin{pmatrix} 1 & 0\\ 0 & 1 \end{pmatrix}, K_1 = \sqrt{\frac{p}{3}} \begin{pmatrix} 0 & 1\\ 1 & 0 \end{pmatrix}$$
\vspace{-20pt}

\vspace{-2pt}
$$\footnotesize K_2 = \sqrt{\frac{p}{3}} \begin{pmatrix} 0 & -i\\ i & 0 \end{pmatrix}, K_3 = \sqrt{\frac{p}{3}} \begin{pmatrix} 1 & 0\\ 0 & -1 \end{pmatrix}$$

\vspace{-8pt}

\section{Our Methodology} \label{sec:methodology}
\vspace{-5pt}
This paper investigates the impact of noise on HyQNNs training.
The HyQNNs used is a simple classifying model using the Iris plant dataset and a variational training model. A detailed overview of our methodology is shown in Fig. \ref{fig:methodology}.


\vspace{-8pt}

\subsection{Requirement Specifications}

\paragraph{Dataset} In our study, we employed the Iris plant dataset \cite{iris_dataset}, which consists of two features. The primary rationale for selecting this dataset was its inherent simplicity, which facilitates straightforward training processes. Additionally, the simplicity of the dataset is adequate for the investigative focus of our study, particularly in examining the behavior and impact of noise gates within this context.

\paragraph{Number of Qubits} For our analysis, we used an Iris plant dataset \cite{iris_dataset} was used which has 2 features and thus a total of 2 qubits were used to contruct hidden quantum layers in HyQNNs architecture. 

\paragraph{Noise and Error Gates}

A total of 5 noise gates were investigated namely; Phase Flip, Bit Flip, Phase Damping, Amplitude Damping, and Depolarizing Channel noise. These gates were chosen for variety in noise sources and their impact on training performance of HyQNNs. 
Each noise gate is tied to a probability of acting on the circuit. The probabilities are chosen from 0.1 to 1.0 in 0.1 increments, having a total of 10 noise probabilities investigated for each noise/error gate. This was chosen in order to show progression of how the learning of an HyQNNs is impacted as probability of noise/error induction steadily increases.

\vspace{-5pt}
\subsection{HyQNNs Architecture}
The HyQNNs architecture used for our analysis typically follows the general architecture as explained in Section \ref{sec:background}. 

\paragraph{Data Encoding}
The first step was to encode the input data into quantum states so that the following quantum layers can process it, as shown in blue-shaded region in HyQNNs architecture part of Fig \ref{fig:methodology}. We used angle encoding (Eq. \ref{eq1}) to encode the input data into quantum state, typically in rotation angles of $RX$ gate. 

\paragraph{Noisy Variational Quantum Circuit}
Once the data is encoded, the next step is to construct VQCs for training. The VQCs are injected separately with all the noise/error gates with different probabilities as described previously. A step-by-step construction of noisy VQC,we have used is discussed below:
\begin{enumerate}
    \item We use two qubit and each qubit is applied with a 3-dimensional arbitrary unitary rotation gates which contains trainable parameters which are tuned during training. 
    
    \item The noise/error gates are then introduced on each qubit.

    \item A CNOT gate is the used to entangle the qubits.

    \item The noise/error gates is again injected on each qubit.
\end{enumerate}

The above steps forms a single layer of the VQC we have used, and this whole layer is repeated 5 times before the measurement. Only the first qubit is measureed in Pauli-Z basis.
It is important to note here that, for noise free training (a bench mark in our comparison), the noise injection steps are omitted and the rest of the steps and specifications remain the same.

\paragraph{Fully Connected Layer}

Once we obtain the qubit measurement results, these outcomes are usually postprocessed through a fully connected classical neural layer, as shown in Fig. \ref{fig:HyQNNs_arch}. 
However, the target problem of this paper is binary classification, as it use to the two classes from the Iris Dataset. Consequently, the measurement results of a single qubit, which generally yield values of -1 (representative of the first class) and 1 (representative of the second class), makes the application of a classical neural network layer at this stage unnecessary. 

Following this, a cost function is defined to evaluate the performance of the network. The specifics of this cost function used in this paper, are detailed in Section \ref{sec:exp_setup}. Based on the values computed by this cost function, the final prediction of the network is then determined.


\begin{figure}[htbp]
    \centering
    \includegraphics[scale=0.45]{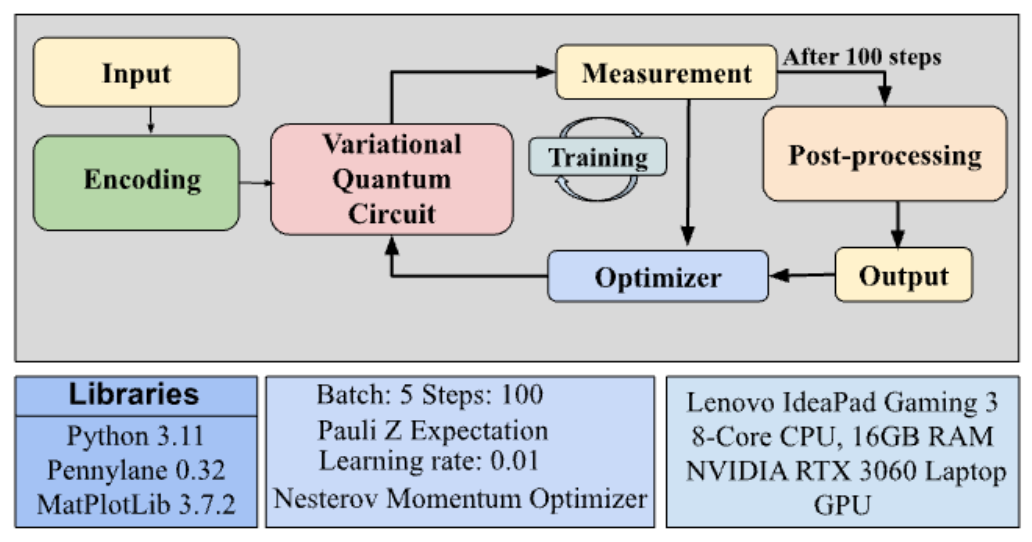}
    \vspace{-8pt}
    \caption{\footnotesize Overview of our Experimental Setup. }
    \label{fig:exp_setup}
\end{figure}
\begin{figure*}[htbp]
    \centering
    \includegraphics[scale=0.7]{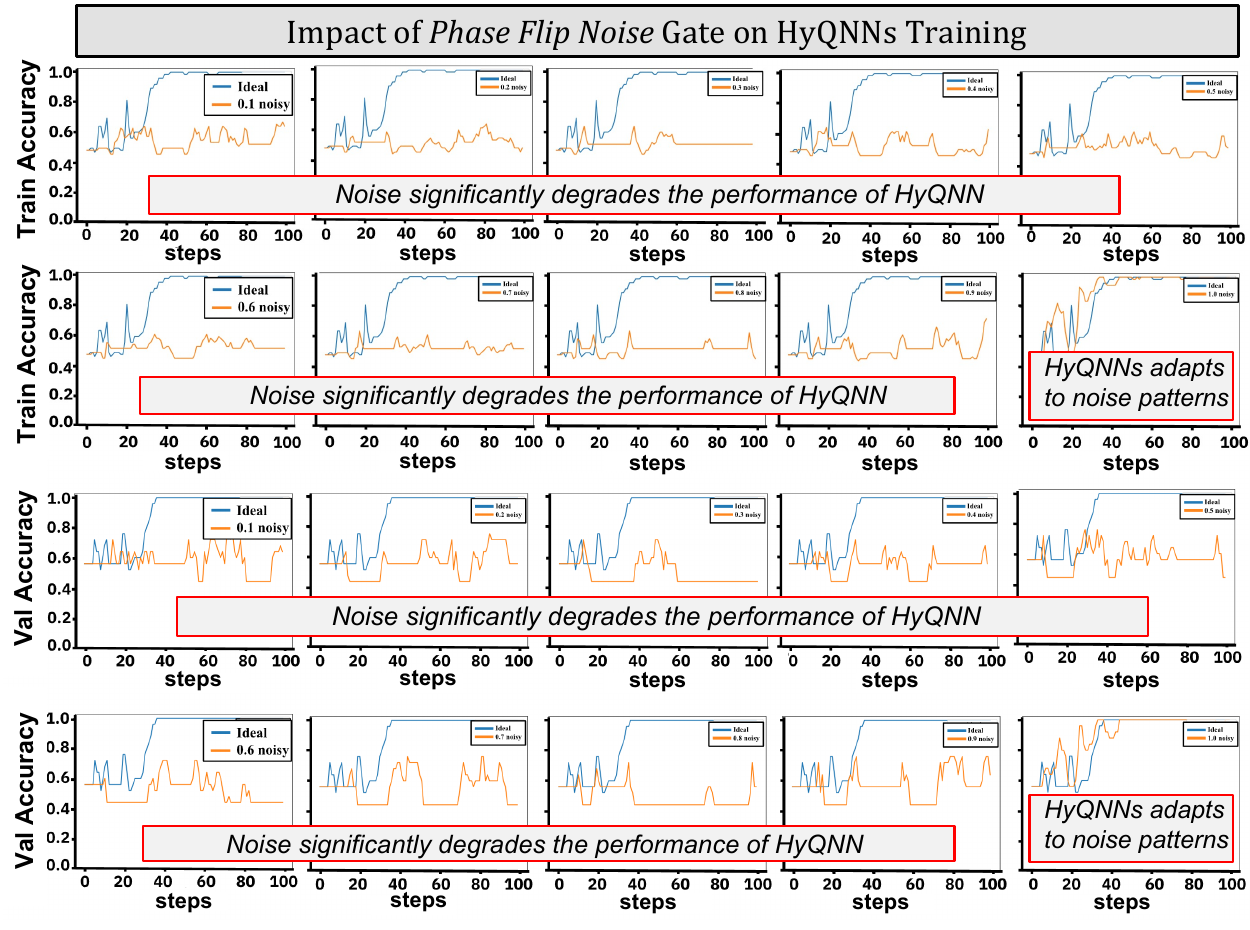}
    \vspace{-12pt}
    \caption{\footnotesize Training and Validation Accuracies of HyQNNs with Phase Flip Noise gates. The top two rows represent training accuracy and bottom two rows represent vaalidation accuracy. The noise probabilities are 0.1 to 0.5 (first and third row, left to right) and 0.6 to 1.0 (second and fourth row, left to right). Blue line represents noise-free training and orange line represents noisy training.
}
    \label{fig:phase_flip_train_val}
\end{figure*}

 \vspace{-20pt}
\subsection{Analysis}

To assess the impact of noise on the HyQNNs, our analysis will focus on key performance metrics: the training and validation accuracies of the HyQNNs. We will conduct a comparative analysis of these accuracies under two distinct conditions - with and without the presence of noise in the quantum circuits. This comparison aims to elucidate how the introduction of noise affects the HyQNNs's performance throughout the training process. Additionally, it will provide insights into the distinct characteristics and effects of different noise gates on the HyQNNs, thereby offering a deeper understanding of their unique impacts on quantum computational accuracy and efficiency.

\vspace{-8pt}

\section{Experimental Setup} \label{sec:exp_setup}
An overview of the experimental setup used to carry out the analysis conducted in this paper is presented in Fig. \ref{fig:exp_setup}. 
The cost function used is defined by the square loss, a simple metric for the distance between the expected value and the predicted value of HyQNNs, described by the following equation, where $y$ is the given data and $h(x)$ is the predicted value.
\vspace{-3pt}
$$C=(y-h(x))^2$$

The HyQNNs are then trained for 100 iterations, with each iteration processing a batch of 5 data points. Initially, the data undergoes a preprocessing stage, which is essential for adapting classical information into a format suitable for quantum processing. This preprocessing involves padding and normalizing the data to align with the requirements of the qubits. Subsequently, the features of the preprocessed data are extracted and encoded into the qubits.

Once the quantum state preparation is complete, this encoded quantum information is fed into the variational quantum circuit. The circuit's operations manipulate the qubit states, and the resulting quantum states are then measured. The outcomes of these measurements are then postprocessed, where the defined cost function evaluates the performance of the HyQNNs.
Nesterov Momentum Optimizer with a learning rate of 0.01 is employed to adjust the parameters of the VQC, aiming to minimize the cost function. 
The decision to use 100 steps is to give the network the opportunity for the noise gates to activate with relative frequency across the probabilities and to analze if the HyQNNs can possibly learn or take into account the error.



\begin{figure*}[htbp]
    \centering
    \includegraphics[scale=0.7]{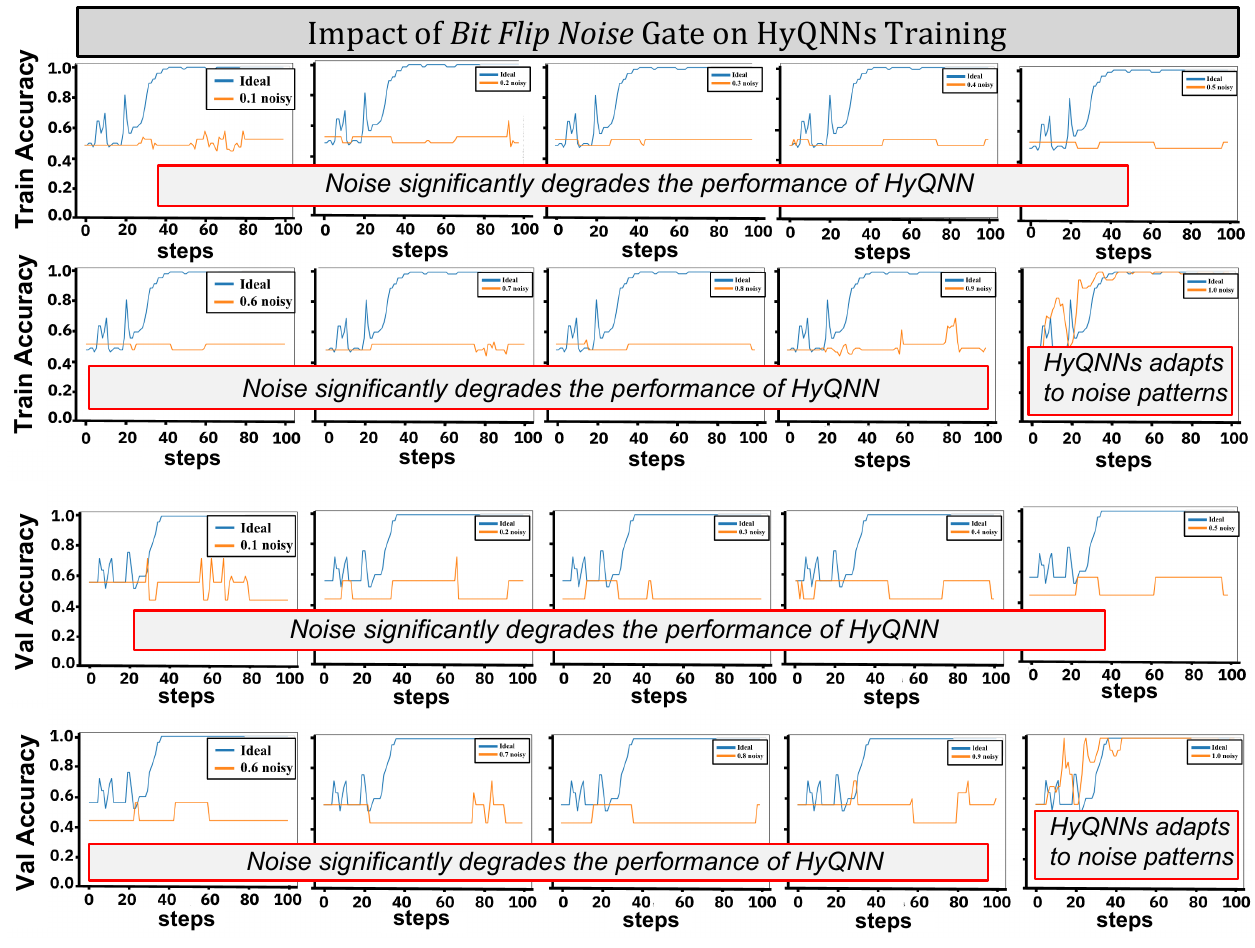}
    \vspace{-12pt}
    \caption{\footnotesize Training and Validation Accuracies of HyQNNs with Bit Flip Noise gates. The top two rows represent training accuracy and bottom two rows represent vaalidation accuracy. The noise probabilities are 0.1 to 0.5 (first and third row, left to right) and 0.6 to 1.0 (second and fourth row, left to right). Blue line represents noise-free training and orange line represents noisy training.
}
    \label{fig:bitFlip_train_val}
\end{figure*}

\vspace{-5pt}

\section{Results and Discussion} \label{sec:results}

\vspace{-5pt}

\subsection{Phase Flip Error}
The introduction of phase flip noise into HyQNNs significantly impacts their performance in terms of training and prediction capabilities. The results are presented in Fig. \ref{fig:phase_flip_train_val}, demonstrates that noise-free training yields higher accuracy compared to training under noisy conditions. 

We observe that when the Phase Flip error probability is set to 1.0, the HyQNNs achieves comparable final accuracies in both training and validation as it does on an ideal, noise-free device. This outcome is attributable to the intrinsic characteristics of the Phase Flip gate. Functioning analogously to a Pauli Z gate, the Phase Flip introduces an error that impacts the qubit's phase without altering its informational content. The principal challenge associated with this gate is the unpredictability of the error, especially at intermediate probabilities ranging from 0.5 to 0.8, where the accuracies for both training and validation are observed to be the lowest.

However, at a probability of 1.0, the Phase Flip error becomes a constant and predictable factor. Under these conditions, the HyQNNs can effectively adapt to this consistent noise pattern during training, as if it is operating in a noise-free environment. This adaptability highlights the resilience of the HyQNNs to certain types of noise, particularly when these noise factors are consistent and can be accounted for during the training phase.

\vspace{-5pt}
\subsection{Bit Flip Error}
A Bit Flip error, represented by the Pauli X gate, preserves certain information while introducing errors. The training results with noise-free and with bit flip noise are shown in Fig. \ref{fig:bitFlip_train_val}. When the probability of a Bit Flip error reaches 1.0, we observe that the HyQNNs performs effectively, similar to that of an ideal, error-free scenario. 
This resilience to Bit Flip errors at high probability is notable; however, it's important to highlight the distinct overall performance impact of Bit Flip errors compared to Phase Flip errors on the HyQNNs.

The performance disparity becomes evident when examining the HyQNNs's accuracy across various probabilities of Bit Flip errors, excluding the 1.0 probability case. Notably, there are noticeable plateaus in accuracy at these probabilities. These plateaus indicate that the HyQNNs faces challenges in adapting its parameters effectively in the presence of Bit Flip errors. Unlike Phase Flip errors, the HyQNNs under Bit Flip conditions struggles not only to enhance accuracy but, in some instances, even experiences a decrease in accuracy.

This differential impact implies that while the HyQNNs can adapt to specific types of errors (like those induced by the Phase Flip gate) by adjusting its parameters, it finds it more challenging to do so with Bit Flip errors. This could be due to the nature of the Bit Flip error, which may affect the quantum states in a manner more disruptive to the HyQNNs's learning process. 
Consequently, this highlights an area of potential improvement in the design and training methodologies of HyQNNs to better handle different types of quantum errors.

\begin{figure*}[htbp]
    \centering
    \includegraphics[scale=0.7]{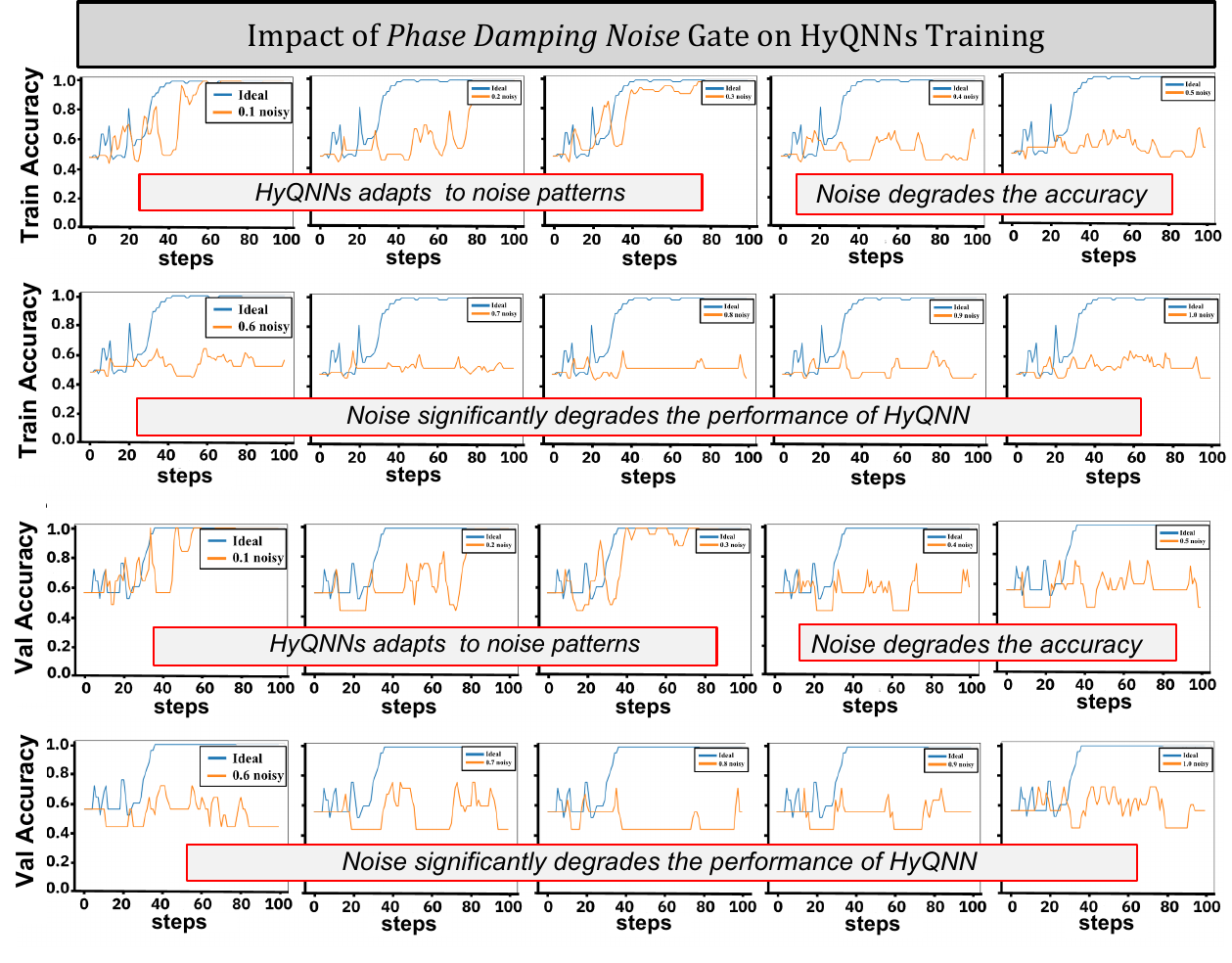}
    \vspace{-10pt}
    \caption{\footnotesize Training and Validation Accuracies of HyQNNs with Phase Damping Noise gates. The top two rows represent training accuracy and bottom two rows represent vaalidation accuracy. The noise probabilities are 0.1 to 0.5 (first and third row, left to right) and 0.6 to 1.0 (second and fourth row, left to right). Blue line represents noise-free training and orange line represents noisy training.
}
    \label{fig:phase_damp_train_val}
\end{figure*}
\vspace{-7pt}
\subsection{Phase Damping}
The Phase Damping gate exerts a significant influence on the quantum information encoded in qubits. In the context of a HyQNNs, its impact on the network's performance can be systematically analyzed by varying the probability of Phase Damping error. The training comparison results of HyQNNs with phase damping noise and noise-free scenarios are shown in Fig. \ref{fig:phase_damp_train_val}
At lower probabilities, ranging from 0.1 to 0.3, the HyQNNs exhibits comparable accuracy to that of noise-free scenario. Although there are some downward spikes in the training curve at these probabilities, however, despite these perturbations, the HyQNNs demonstrates a robust ability to achieve high accuracy levels, suggesting a degree of resilience to low levels of Phase Damping noise.

As the probability of Phase Damping error increases beyond this low range, the accuracy declines become more pronounced. This is characterized by more frequent downward spikes and a general trend of reduced accuracy in both training and validation phases. Notably, at a probability of 0.8, there is an extended period of diminished accuracy, which is manifested as a prolonged, deep plateau. This indicates a significant degradation in the QNN's performance.

The underlying reason for this observed behavior can be attributed to the nature of the Phase Damping gate. This type of noise gate specifically alters the phase information in the qubits, which is critical for the QNN's processing and learning capabilities. As the probability of Phase Damping error increases, the QNN's ability to effectively train and iteratively refine its parameters is increasingly compromised. This leads to a marked decline in performance, reflecting the network's diminished capacity to accurately process and learn from the quantum data.

Thus, the impact of Phase Damping errors on a HyQNNs is not only a matter of reduced accuracy but also indicates a fundamental limitation in the network's ability to cope with certain types of quantum noise, especially at higher probabilities. This highlights the need for developing more robust quantum error mitigation strategies in HyQNN's design.

\begin{figure*}[htbp]
    \centering
    \includegraphics[scale=0.7]{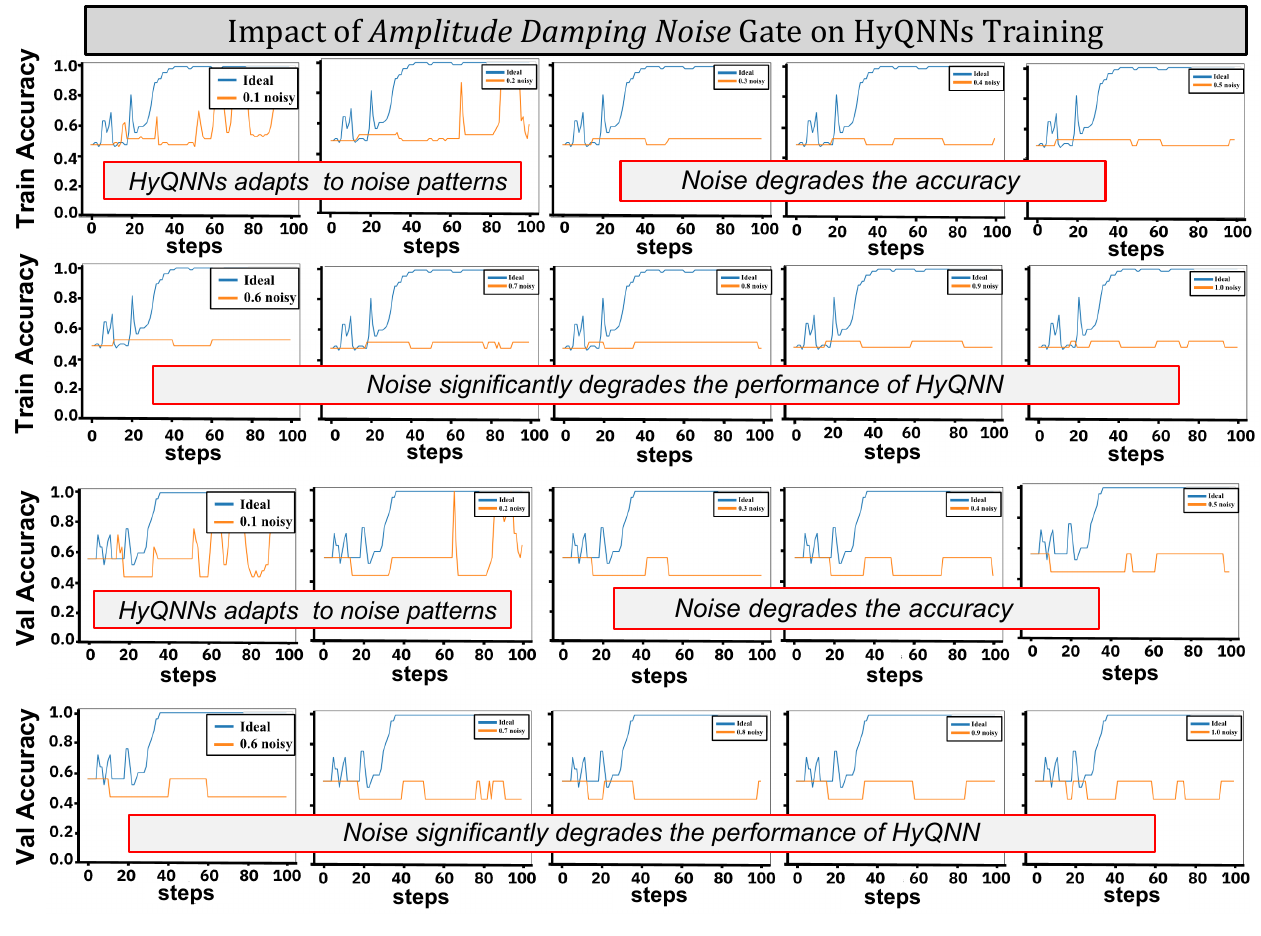}
    \vspace{-18pt}
    \caption{\footnotesize Training and Validation Accuracies of HyQNNs with Amplitude Damping Noise gates. The top two rows represent training accuracy and bottom two rows represent vaalidation accuracy. The noise probabilities are 0.1 to 0.5 (first and third row, left to right) and 0.6 to 1.0 (second and fourth row, left to right). Blue line represents noise-free training and orange line represents noisy training.
}
    \label{fig:amp_damp_train_val}
\end{figure*}

\vspace{-9pt}
\subsection{Amplitude Damping}

Amplitude Damping, similar to that of Phase Damping, results in the loss of quantum information, which has consequential effects on the performance of HyQNNs. The impact of Amplitude Damping on a HyQNNs can be systematically evaluated by observing its training and validation accuracies across various probabilities, as shown in Fig. \ref{fig:amp_damp_train_val}.

At lower probabilities of Amplitude Damping, specifically at 0.1 and 0.2, the HyQNNs exhibit some training potential. However, a notable distinction emerges when comparing the effects of Amplitude Damping with those of Phase Damping. The HyQNNs exhibits a prolonged training duration to achieve high accuracies under the influence of Amplitude Damping. 

The influence of Amplitude Damping extends to higher probabilities as well. In these scenarios, the performance of the HyQNNs is largely characterized by extended periods of low accuracy, denoted as long plateaus. This pattern suggests that the HyQNNs struggles to adapt its parameters effectively in the presence of higher levels of Amplitude Damping noise.

The underlying mechanism for this behavior can be attributed to the nature of Amplitude Damping, which primarily affects the amplitude of the quantum state. Such damping leads to a reduction in the superposition states that are crucial for quantum computation. As a result, the HyQNNs's ability to process and learn from quantum data is substantially hindered, particularly at higher probabilities of error.


\begin{figure*}[htbp]
    \centering
    \includegraphics[scale=0.7]{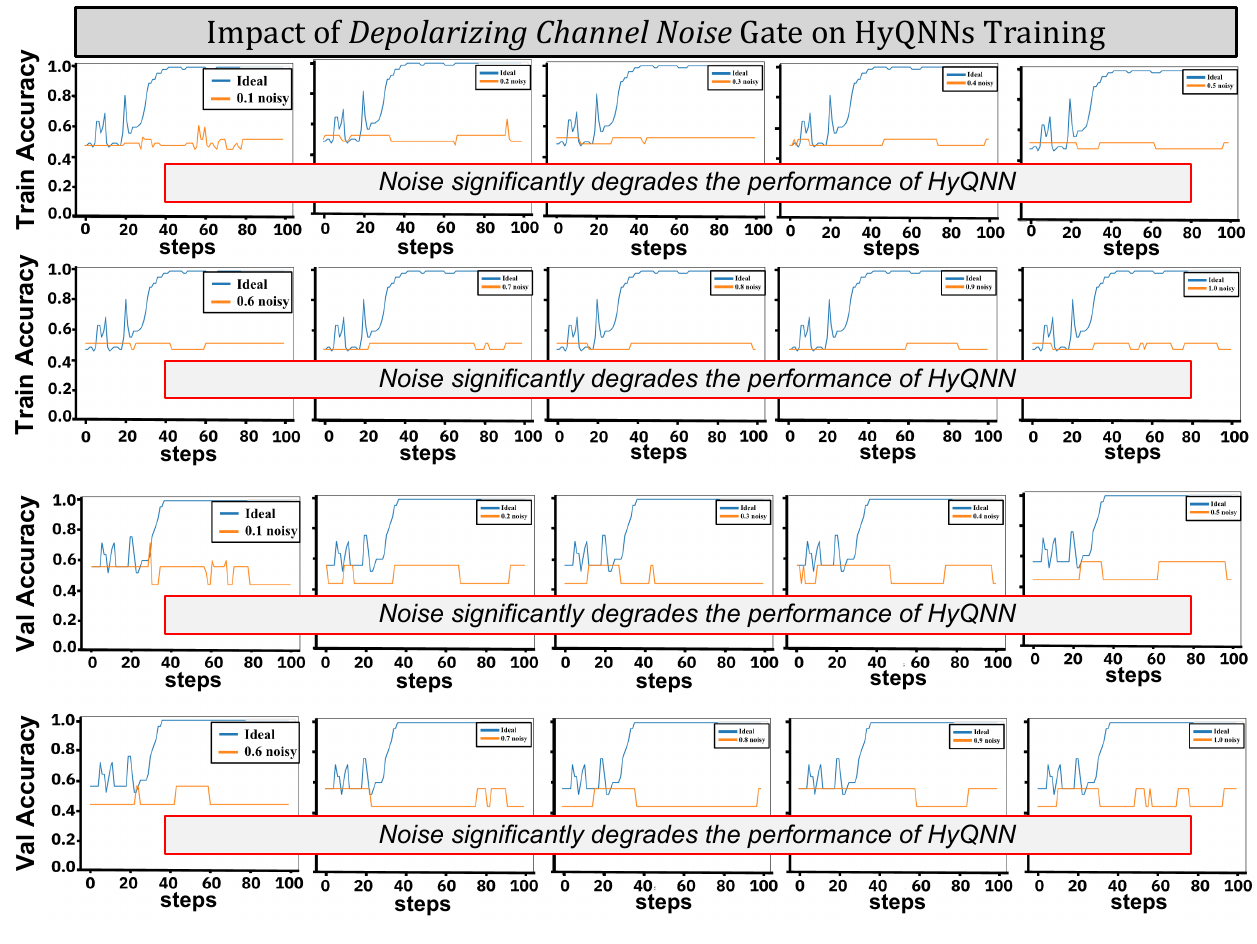}
    \vspace{-18pt}
    \caption{\footnotesize Training and Validation Accuracies of HyQNNs with Depolarizing Channel Noise gates. The top two rows represent training accuracy and bottom two rows represent vaalidation accuracy. The noise probabilities are 0.1 to 0.5 (first and third row, left to right) and 0.6 to 1.0 (second and fourth row, left to right). Blue line represents noise-free training and orange line represents noisy training.
}
    \label{fig:depolar_channel_train_val}
\end{figure*}

\vspace{-4pt}
\subsection{Depolarizing channel}
The Depolarizing Channel, among various noise gates experimented in this paper, exhibits the most detrimental impact on the performance of HyQNNs across a range of probabilities. This is evident from the consistently low trends observed in both training and validation accuracies under the influence of the Depolarizing Channel, as shown in Fig. \ref{fig:depolar_channel_train_val}.

A key characteristic of the HyQNNs's performance in the presence of Depolarizing noise is the limited and transient nature of any training improvements. 

Notably, slight improvements in training and validation accuracies are primarily observed at the lower probability of 0.1, with marginal indications of this trend at a probability of 0.2. 
This observation implies that at lower levels of randomness, the HyQNNs exhibits a certain degree of resilience to Depolarizing noise, however, this resilience is significantly limited and does not translate into substantial performance gains.

The underlying reason for the pronounced negative impact of the Depolarizing Channel on HyQNNs can be attributed to the nature of depolarizing noise. This type of noise introduces randomness into the quantum state, leading to a loss of coherence and information critical for quantum computation. As a result, the HyQNNs's ability to effectively process and learn from quantum data is severely compromised, particularly at higher probabilities of error.

The performance of HyQNNs under the Depolarizing Channel is significantly hindered across all probabilities. This highlights the critical need for advanced error mitigation techniques in quantum computing, particularly in addressing the challenges posed by depolarizing noise environments.

\vspace{-8pt}

\begin{table}[htbp]
\caption{\footnotesize Quantitative Summary of all Noise Gates at Different Probability Values. The $\checkmark$ shows the cases where HyQNNs shows the potential to adapt or learn the noise patterns and $\times$ shows the cases where HyQNNs shows almost no potential of training and succumbs to adverse affects of noise.}
\label{tab:table}
\footnotesize
\begin{tabular}{|c|c|l|l|l|l|l|l|l|l|l|l|}
\hline
\multirow{2}{*}{\makecell{\textbf{Noise} \\ \textbf{Gates}}} & \multicolumn{10}{c|}{\textbf{Probabilities}} \\ \cline{2-11} 
                             & \textbf{0.1} & \textbf{0.2} & \textbf{0.3} & \textbf{0.4} &\textbf{ 0.5} & \textbf{0.6} & \textbf{0.7} & \textbf{0.8} & \textbf{0.9} & \textbf{1.0} \\ \hline
Phase Flip                   & $\times $& $\times$ & $\times$ & $\times$ & $\times$ & $\times$ & $\times$ & $\times$ & $\times$ & $$\checkmark$$ \\ \hline
Bit Flip                     & $\times$ & $\times$ & $\times$ & $\times$ & $\times$ & $\times$ & $\times$ & $\times$ & $\times$ & $\checkmark$ \\ 
\hline

\makecell{Phase \\ Damping}           & $\checkmark$    & $\checkmark$     & $\checkmark$     &$\times$     &$\times$     & $\times$    & $\times$    & $\times$    & $\times$    & $\times$    \\ \hline

\makecell{Amplitude \\ Damping}         &$\checkmark$    &$\checkmark$     &$\times$     &$\times$     & $\times$    &$\times$     & $\times$    &  $\times$   & $\times$    & $\times$    \\ \hline
\makecell{Depolarizing \\ Channel}      &  $\times$   & $\times$    & $\times$    & $\times$    & $\times$    & $\times$    & $\times$    & $\times$    &$\times$     &$\times$     \\ \hline
\end{tabular}

\end{table}


\section{Conclusion} \label{sec:conclusion}
In this paper, we conducted a comprehensive quantitative analysis of how quantum noise gates affect the performance of Hybrid Quantum Neural Networks (HyQNNs), a summary of which is presented in Table \ref{tab:table}. Our analysis revealed a complex interplay between quantum noise and HyQNNs capabilities, highlighting both strengths and limitations. We found that Phase Flip gates, akin to Pauli Z gates, exhibited a unique resilience at a probability of 1.0, allowing HyQNNs to adapt to consistent noise patterns. However, this adaptability varied, with performance decreasing at intermediate error rates. Similarly, Bit Flip errors, represented by Pauli X gates, led to reduced accuracy at lower probabilities but showed improved HyQNNs performance at higher probabilities. Intricacies were further observed with Phase Damping and Amplitude Damping gates. While HyQNNs showed resilience to Phase Damping at lower probabilities, they experienced a notable decline in performance at higher levels. Amplitude Damping posed a significant challenge, limiting training efficiency and resulting in performance plateaus at higher probabilities. The most substantial and adverse effects were under the Depolarizing Channel, where HyQNNs shows no learning capabilities, irrespective of the probability of Depolarizing channel noise gate. 

The underlying mechanisms behind these observations lie in the nature of each noise gate. Phase Flip, Bit Flip, Phase Damping, Amplitude Damping, and the Depolarizing Channel each introduce distinct challenges, from altering phase information to disrupting quantum coherence. Our findings underscore the importance of developing tailored quantum error correction and mitigation strategies to fully harness the potential of HyQNNs, considering the distinct challenges posed by each type of quantum noise.


\vspace{-5pt}
\begin{spacing}{0.85}
    \section*{Acknowledgements}
\vspace{-0.14cm}
This work was supported in part by the NYUAD Center for Quantum and Topological Systems (CQTS), funded by Tamkeen under the NYUAD Research Institute grant CG008.
\end{spacing}

\vspace{-4pt}
\begin{spacing}{0.75}
\bibliographystyle{ieeetr}
\bibliography{references.bib} 
\end{spacing}

\end{document}